# Fracture Characterization of Bioinspired Irregular Network Reinforced Composites

*Chelsea Fox[1], Tommaso Magrini[2*] and Chiara Daraio[1*]*

[1] Division of Engineering and Applied Science, California Institute of Technology, Pasadena, CA 91125, USA

[2] Department of Mechanical Engineering, Eindhoven University of Technology, 5600MB Eindhoven, The Netherlands

[*] Email: t.magrini@tue.nl, daraio@caltech.edu

**Abstract**

The mechanical behavior of composite materials is significantly influenced by their structure and constituent materials. One emerging class of composite materials is irregular network reinforced composites (NRC's), whose reinforcing phase is generated by a stochastic algorithm. Although design of the reinforcing phase network offers tailorable control over both the global mechanical properties, like stiffness and strength, and the local properties, like fracture nucleation and propagation, the fracture properties of irregular NRC's has not yet been fully characterized. This is because both the irregular reinforcing structure and choice of matrix phase material significantly affect the fracture response, often resulting in diffuse damage, associated with multiple crack nucleation locations. Here, we propose irregular polymer NRC's whose matrix phase has a similar stiffness but half the strength of the reinforcing phase, which allows the structure of the reinforcing phase to control the fracture response, while still forming and maintaining a primary crack. Across a range of network coordination numbers, we obtain J-integral and R-curve measurements, and we determine that low coordination polymer NRC's primarily dissipate fracture energy through plastic zone formation, while high coordination polymer NRC's primarily dissipate energy through crack extension. Finally, we determine that there are two critical length scales to characterize and tailor the fracture response of the composites across the coordination numbers: (i) the size of the plastic zone, and (ii) the size and geometry of the structural features, defined as the areas enclosed by the reinforcing network.

**Keywords**

Fracture mechanics, composite materials, irregular materials, bioinspired materials

## 1. Introduction

Architected materials often draw inspiration from the structure of biological materials to achieve desirable properties such as high stiffness and strength (Fox et al., 2024; Jung et al., 2024; Zhang et al., 2021), good energy dissipation (Fox et al., 2024; Gu et al., 2017b; Li et al., 2024; Zhuang et al., 2023), and high fracture toughness (Dai et al., 2023; Dimas et al., 2013; Gu et al., 2017a; Jia and Wang, 2019; Jung et al., 2024; Magrini et al., 2023, 2022, 2019). These bioinspired materials also often feature two or more phases (Espinosa et al., 2009; Gu et al., 2017a; Li et al., 2024; Libonati et al., 2019, 2014; Magrini et al., 2019), as biological materials are typically composite materials, such as bone (Keaveny et al., 2003; Launey et al., 2010; Magrini et al., 2021; Whitehouse, 1974) or nacre (Ackson et al., 1988; Meyers et al., 2008), with mechanical properties superior to that of their constitutive elements. Although most bioinspired composite materials feature repeating structural patterns (Gao et al., 2023; Gu et al., 2016; Li et al., 2022; Magrini et al., 2022, 2019; Studart, 2012), many biological materials have a non-periodic structure (Jentzsch et al., 2024, 2022; Metzler et al., 2019; Studart, 2012), indicating that evolution may favor irregularity as a way to optimize function (Metzler et al., 2019).

To explore the role of irregularity, virtual growth algorithms (VGA) have been developed to imitate the stochastic growth process of biological material structures (Jia et al., 2024a; Liu et al., 2022). These algorithms generate a continuous irregular network by assembling a set of building blocks according to local connectivity rules, and studies have been conducted to determine the relationship between the structure and function of these networks (Jia et al., 2024a, 2024b; Liu et al., 2022), including for polymer composite materials (Fox et al., 2024; Magrini et al., 2023), whose reinforcing phase is generated by the VGA. It has been shown that controlling the average network coordination (Thorpe and Duxbury, 2002; Thorpe, 1983), defined as the number of branches in each building block, primarily influences the stiffness and strength of the materials (Fox et al., 2024; Jia et al., 2024a; Magrini et al., 2023; Zhou et al., 2024), while controlling the connectivity rules determines the formation of specific structural features, defined as the areas enclosed by the reinforcing network (Fox et al., 2024; Magrini et al., 2023).

However, despite the large design space offered by the choice of building blocks and their connectivity rules, polymer composites reinforced by stiff irregular networks with a compliant matrix display the same fracture behavior (Magrini et al., 2023). Fracture begins with the nucleation of voids in the compliant matrix, similar to what is observed in ductile metals (Anderson, 2017), followed by bridging of the reinforcing phase, similar to what occurs in fiber reinforced composites (Chen et al., 1995; Khan, 2019; Nemat-Nasser and Hori, 1987; Sakai et al., 1991). During fracture bridging, the most extensible portions of the reinforcing phase deform in the direction of the applied load, until they undergo local yielding followed by sequential strut failure (Magrini et al., 2023). The sequential failure of the reinforcing phase causes the coalescence of the large-scale voids, leading to the overall loss of structural integrity in the composites (Magrini et al., 2023).

Previous efforts have shown how to nucleate and guide the fracture path (Magrini et al., 2023), but the fracture properties of irregular network reinforced composites have not yet been fully characterized. To better understand the role of the reinforcing phase during fracture, we generate centimeter-scale polymer composite materials (with millimeter scale features), whose matrix phase has a similar stiffness but half the strength of the reinforcing phase. This prevents the diffuse nucleation of voids across the sample and maintains a primary crack, making fracture characterization measurements possible, while still allowing the reinforcing phase to significantly influence the fracture response. Across a range of coordinations from 2.3 to 2.8, we determine that there are two critical length scales necessary to understand and describe the fracture response of the composites: (i) the plastic zone size, and (ii) the structural feature size (and geometry). We then show how to tailor the tradeoff between plastic zone size and primary crack extension using the coordination number and its effect on the structural feature populations ahead of the crack tip.

## 2. Material and Methods

We generate the irregular reinforcing phase of the polymer composite samples using a VGA (Liu et al., 2022). This computer-assisted material design tool generates the irregular samples on a square grid from a set of three tile types, with coordination numbers of either two, for (-) tiles and (L) tiles, or three, for (T) tiles (Figure 1a). Using tile frequency hints and a set of connectivity rules (Figure 1b), the VGA assembles the tiles into continuous irregular networks (Figure 1c), with a characteristic length defined as one tile, which is equivalent to the smallest possible structural feature that can be formed, with a size of 1 mm (Figure 1c). We generate samples with an average coordination that ranges from 2.3 to 2.8 (Figure 1d), which spans the transition from floppy to rigid behavior while maintaining percolation ("Rigidity Theory and Applications," 2002; Thorpe, 1983). The 2.3 coordination samples are composed of 5% (-) tiles, 65% (L) tiles and 30% (T) tiles, the 2.5 coordination samples are composed of 5% (-) tiles, 45% (L) tiles and 50% (T) tiles, and the 2.8 coordination samples are composed of 10% (-) tiles, 10% (L) tiles and 80% (T) tiles, with equally represented two-fold ((-) tiles) or four-fold rotations ((L) and (T) tiles). To determine the sample size needed for testing, we determine the minimum number of tiles required for the network to be representative of the coordination number's population of structural features by finding an exponential fit and looking for when the exponential fit parameters converge as the sample size increases (Figure 1j, 1k). We find that samples of 30x30 tiles are sufficiently representative (Figure 1f-i), but we also note that networks with higher coordination reach a plateau more quickly than lower coordination ones, as the structural feature size depends on the coordination number (Figure 1k).

To fabricate our polymer composite samples for single edge notch plate tension tests, we first choose the volume fraction of the reinforcing phase in each tile, with values of 20% for 2-coordination tiles and 30% for 3-coordination tiles, corresponding to a branch width of 0.2 mm. We then add a matrix phase with a similar stiffness (~1 GPa) (Dizon et al., 2018; Jia and Wang, 2019; Magrini et al., 2023; Slesarenko and Rudykh, 2018), but half the strength of the reinforcing phase (Figure 1e) to fill the remaining space in each tile. We use a Polyjet printer (Stratasys Objet500 Connex3) to additively manufacture the composite materials using a commercially available photoresin (VeroWhite Polyjet Resin, Stratasys) for the reinforcing network, and a compatible photoresin with lower strength (Stratasys Grey60 Polyjet Resin) for the matrix (Figure 1e). Furthermore, the composite materials have a strong adhesion between phases, key to avoid delamination during fracture testing.

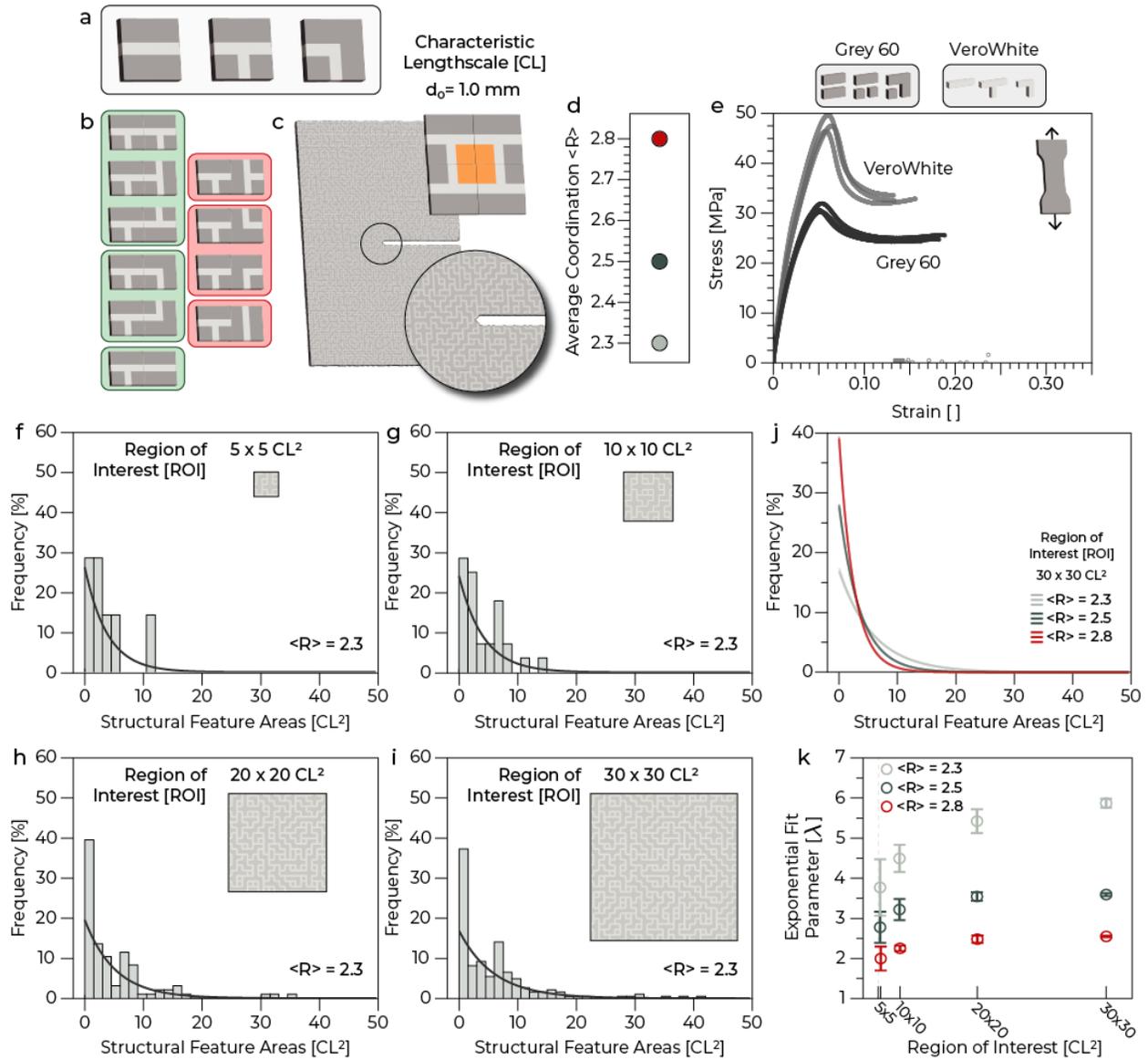

**Figure 1.** Material and structural characterization. (a) VGA tiles and coordination numbers. (b) VGA connectivity rules, green is allowed, red is not allowed. (c) Fracture samples with characteristic length scale. (d) Average coordination number. (e) Constitutive uniaxial stress-strain

plot for reinforcing and matrix phase materials. (f-i) Frequency of 2.3 coordination structural feature areas for 5x5, 10x10, 20x20 and 30x30 tiles. (j) Exponential fits of structural feature areas for 2.3, 2.5 and 2.8 coordination samples with 30x30 tiles. (k) Exponential fit parameters for increasing sizes of sample regions of interest.

## 3. Results and Discussion

### 3.1. Single Edge Notch Plate Tension Fracture Tests

We conduct single edge notch plate tension fracture tests on composite samples of 9x6 cm, corresponding to 90x60 tiles (or characteristic lengths), with a thickness of 1mm and an initial crack of length 3 cm, which we sharpen with a razor blade prior to testing. Grip areas of height 3 cm, manufactured from the same photoresin as the reinforcing phase, are added during printing to the top and bottom of the sample. We use an Instron E3000 (Instron, USA) equipped with a 5kN load cell to apply a tensile load at a rate of 2 mm/min and we test three different samples for each coordination, repeating one of these samples three times to verify that the same structure fractures self-consistently. We compare our results to samples exclusively composed of the matrix and reinforcing phases (Figure 2a,e), and similar to the bulk materials, the composite samples maintain one primary crack, which initiates and propagates through both phases as the tensile loading is applied. This is the consequence of the relatively low mismatch in mechanical properties between the reinforcing and matrix phase polymers, which prevents crack arrest at the interface (Kolednik et al., 2014). Nonetheless, despite having similar volume fractions of reinforcing phase across the coordination numbers, with values of 23%, 25% and 28% for 2.3, 2.5, and 2.8 coordinations, respectively, the force-displacement curves (Figure 2b-d) display significant variations. Among composite samples, 2.3 coordination samples fail at the highest global strain, while 2.8 coordination samples fail at the lowest global strain (Figure 2b,d). The measured strain-to-failure values are in between the recorded strain-to-failure values for the bulk materials (Figure 2a,e). To explain these variations, we track the strain fields ahead of the crack tip using 2D digital image correlation. We spray paint the samples with matte white paint and apply matte black speckles of diameter 0.1-0.3 mm and then use VIC 2D (Correlated Solutions, USA) to analyze the Lagrangian strain fields. Unlike previous studies featuring composites with a soft elastomeric matrix, which are prone to multiple void nucleation sites and diffuse damage across the entire sample (Magrini et al., 2023) (Figure SI 1), we observe the formation of a single plastic zone across all samples. Assuming that plastic yielding begins when the maximum principal strain value exceeds the yield strain from the constitutive uniaxial stress-strain data, we apply a threshold and track the size of the plastic zone with respect to the sequential events: yield, crack initiation, and sample failure (Figure 2f-j). Most importantly, we note that the plastic zone in all samples develops to be over one order of magnitude larger than the structural features.

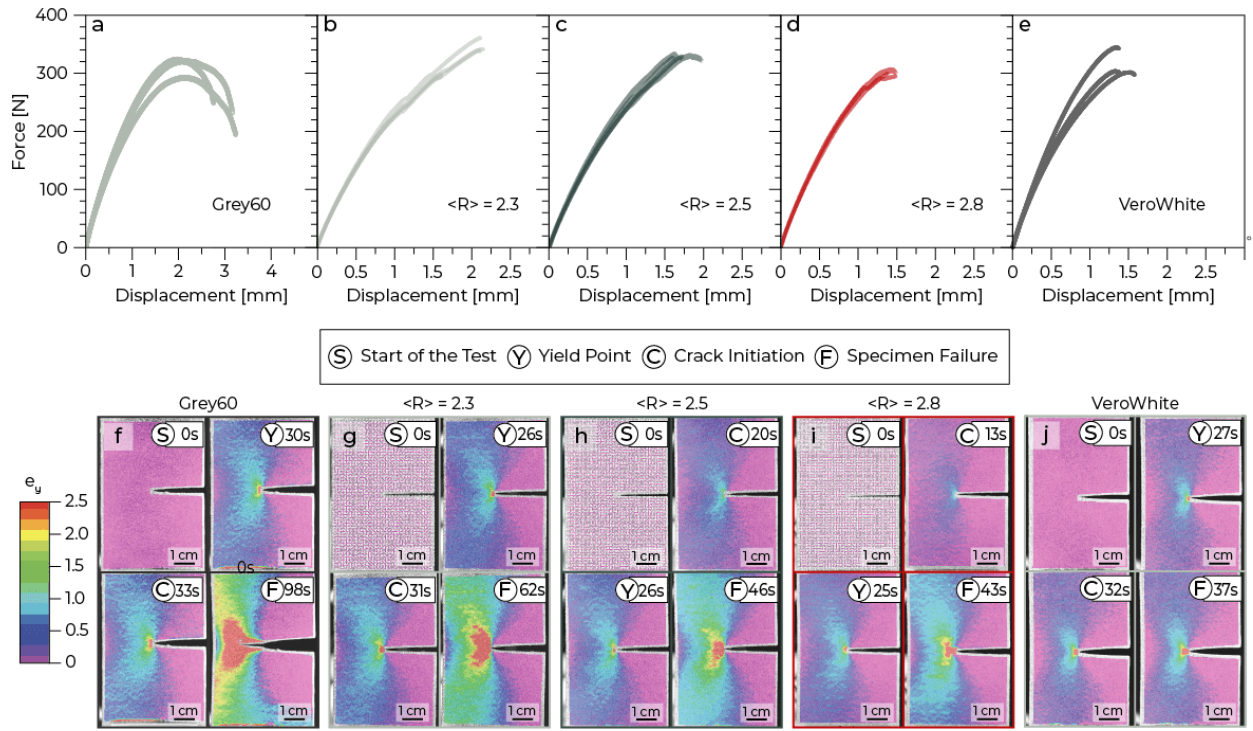

**Figure 2.** Mechanical characterization and DIC images of plastic zone growth for bulk material and composite samples. (a-e) Force-displacement plots for bulk and composite samples. (f-j) 2D DIC strain maps of bulk and composite samples with plasticity threshold in red and time stamps.

### 3.2. Fracture Behavior of the Composites: Plastic Zone Formation and Crack Extension

Although the plastic zone begins to develop at approximately the same global displacement in all samples, we observe that the maximum size of the plastic zone reached by each coordination increases non-linearly with coordination number (Figure 3a, SI 2). By tracking the amount of crack extension with respect to the global displacement applied, using the image processing software, FIJI (Schneider et al., 2012), we also observe that lower coordination samples have less crack extension despite reaching a higher global strain. Indeed, the 2.3 coordination samples do not exceed $1.4 \pm 0.2$ mm of crack extension, while 2.8 coordination samples reach up to $3.5 \pm 0.8$ mm of crack extension before their sudden failure (Figure 3b). This tradeoff between plastic zone size, crack extension, and global strain-to-failure is also reflected in the J-integrals for each sample. We define the J-integral for an edge cracked sample as:

$$J = \frac{\eta U}{Bb} \tag{1}$$

where $\eta$ is a dimensionless constant, $U$ is the area under the force displacement curve, $B$ is the sample thickness, and $b$ is the uncracked length(Anderson, 2017) (Figure 3b, inset). The J-integral can also be written as a sum of its elastic and plastic components:

$$J = J_e + J_p \tag{2}$$

$$J_e = \frac{\eta_e U_e}{Bb} \tag{3}$$

$$J_p = \frac{\eta_p U_p}{Bb} \tag{4}$$

where *e* and *p* refer to the elastic and plastic components, respectively (Anderson, 2017). By extracting the elastic and plastic components of the J-integral incrementally (Figure SI 3), we observe the tradeoff between recoverable and irrecoverable deformation within the samples, as the contribution to the normalized elastic J-integral increases as the coordination number increases, while the contribution to the normalized plastic J-integral decreases as the coordination number increases (Figure 3c).

We then obtain the R-curves for each coordination by plotting the normalized J-integrals as a function of crack extension (Figure 3d). Regardless of their coordination, all samples exhibit a rising trend, although the shape of each R-curve varies significantly across the coordination numbers (Figure 3d). The 2.3 coordination samples have a higher critical J-integral value, $J_c$, at which the crack begins to grow, while the higher coordination samples show an average decrease in the critical $J_c$ of 32.6% and 61.2% for 2.5 and 2.8 coordinations, respectively. After initiation, the 2.3 coordination samples all exhibit subcritical crack extension of a few characteristic lengths before reaching a saturation point, during which the crack arrests and the plastic zone grows until the sample reaches a maximum J-integral value, $J_m$, and the sample fails suddenly (Figure 3d, left). In contrast, the 2.5 coordination samples all exhibit simultaneous crack extension and plastic zone growth before sample failure, reaching a 16.1% lower average $J_m$ than the 2.3 coordination samples (Figure 3d, center). However, unlike the lower coordinations, the 2.8 coordination samples reach a saturation point in plastic zone size, allowing for several characteristic lengths of crack extension before sample failure, and reaching the lowest average $J_m$, 29.4% lower than the 2.3 samples (Figure 3d, right). We then compare the composite materials with the bulk materials of the reinforcing and matrix phases. The bulk material of the matrix phase displays a larger plastic zone area than the 2.3 coordination samples and has a constantly rising R-curve with simultaneous crack extension. In contrast, the bulk material of the reinforcing phase displays a smaller plastic zone area than the 2.8 coordination samples, with a very short R-curve, as the sample fails at very low global strain with little crack extension (Figure SI 4).

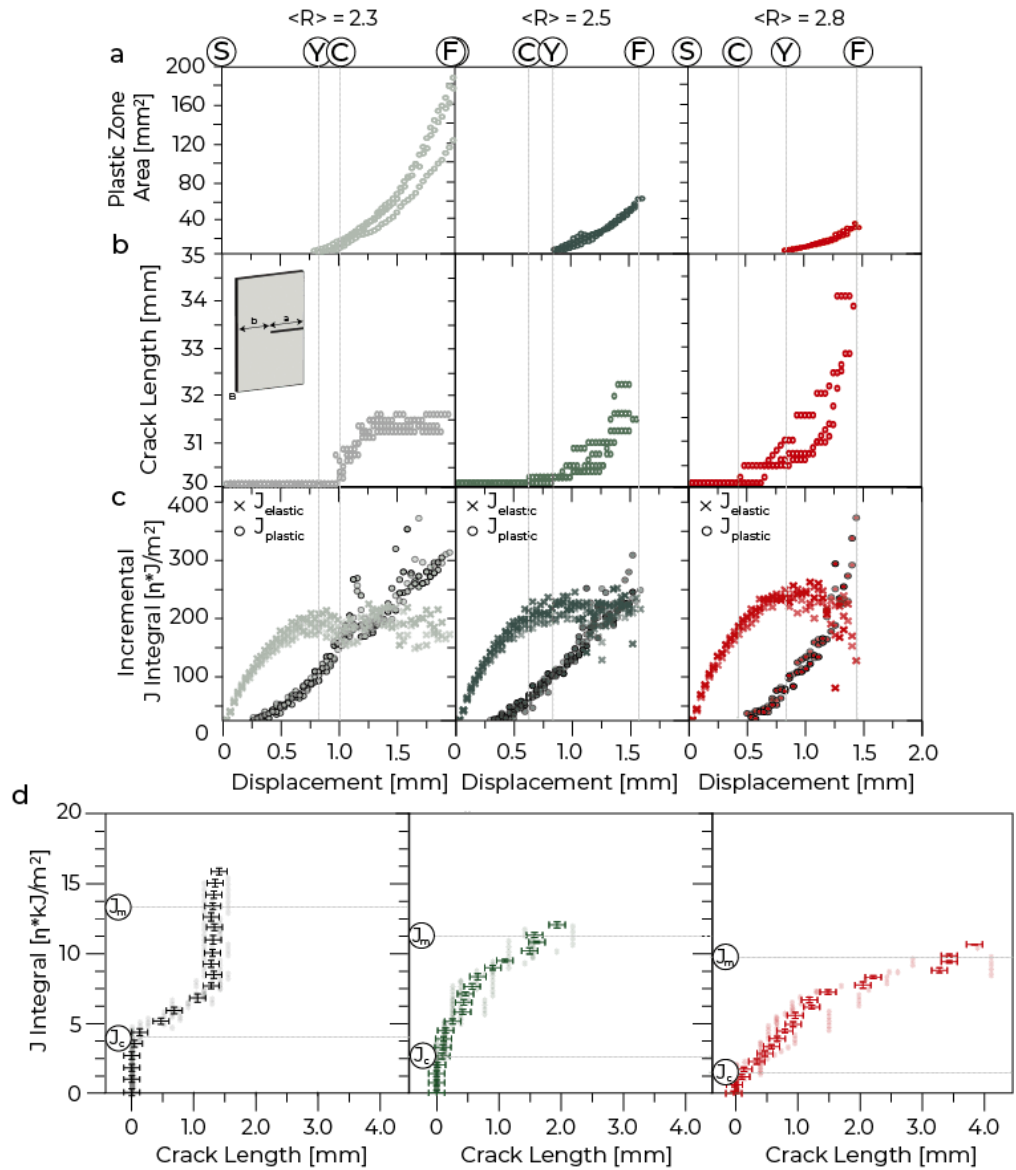

**Figure 3.** Plastic zone size, crack extension length, J-integrals and R-curves for composite material samples. S corresponds to start of test, Y corresponds to yielding, C corresponds to crack initiation, F corresponds to failure. (a) Plastic zone area as a function of displacement for 2.3, 2.5, and 2.8

coordination samples. (b) Crack extension length as a function of displacement for 2.3, 2.5, and 2.8 coordination samples, with inset showing single edge notch plate tension sample dimensions. (c) Incremental elastic J-integrals ($\eta_e$ normalized) and plastic J-integrals ($\eta_p$ normalized) as a function of displacement for 2.3, 2.5, and 2.8 coordination samples. (d) R-curves for 2.3, 2.5, and 2.8 coordination samples.

### 3.3. Effect of Structural Features

Although the plastic zone size is the critical length scale to understand and describe the global fracture behavior (R-curves) of the samples, it is also necessary to consider a second length scale: the relative size and distribution of the structural features. Structural features are composed of groups of tiles which form different populations as a function of coordination number, resulting in plastic zone shape variations across the coordination numbers (Figure 4a). Lower coordination numbers form more polydisperse structural features with a wide range of sizes, from one characteristic length squared up to tens of characteristic lengths squared, resulting in more extensible composite materials, while higher coordination numbers form more monodisperse structural features that are smaller, resulting in stiffer composite materials (Figure 4a, SI 5).

In order to characterize the structural features, we describe their size in terms of their bridge lengths, defined as the distances between (T) tiles (Figure 4b), and we note that the bridge lengths for each coordination are on the same order of magnitude as the characteristic length of 1 mm, which is at least one order of magnitude lower than that of the plastic zones (Figure 4c). It is this relative size ratio, coupled with the reinforcing phase's higher strength, that results in a deformation which not only follows the shape of the local individual structural features ahead of the crack tip, but also globally the characteristic plastic zone shape from linear elastic fracture mechanics (Anderson, 2017) (Figure 4a).

Given that the strain field immediately ahead of the crack tip follows the shape of the local structural features (Figure 4a), it is important to examine the orientation and geometry of these local features (Figure 4d,e) to explain why higher coordination samples have greater crack extension. In order to generate higher coordination numbers, because they are stochastically generated on a square grid, samples are forced to form many smaller linear structural features that are aligned in the direction of the primary crack, at 0° (Figure 4f). These linear features (Figure 4e, red) can be more easily split apart, allowing for more crack extension at lower global strains. In contrast, lower coordination samples have fewer reinforcing interfaces to pass through than high coordination samples, and instead form many larger complex structural features that are at an angle to the primary crack, greater than 0° (Figure 4f). It is more difficult for the crack to proceed through these extensible diagonal features (Figure 4e, gray, green), resulting in crack arrest and allowing the sample to instead dissipate energy through the formation of a large plastic zone.

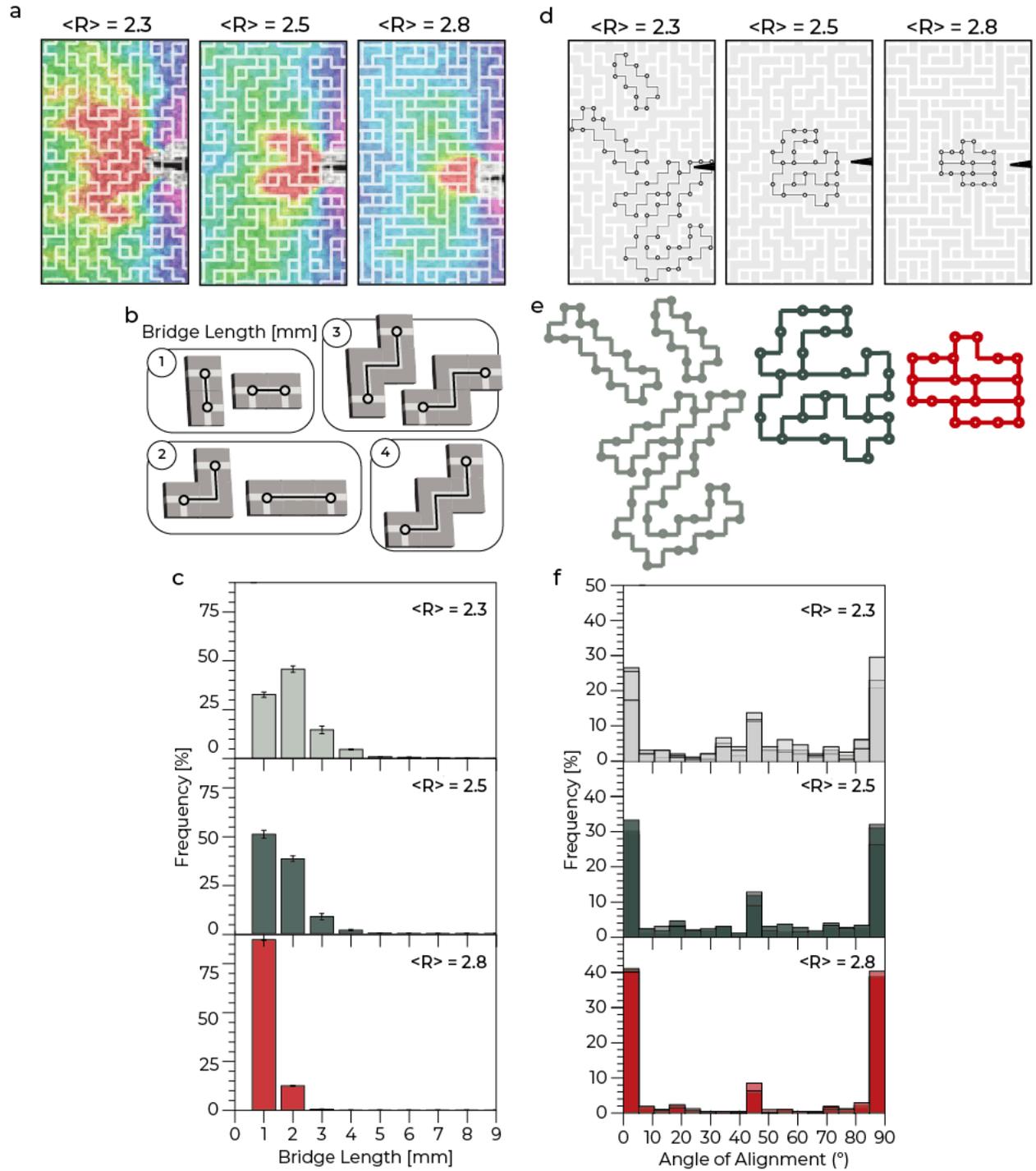

**Figure 4.** Structural feature analysis. (a) Zoomed images of maximum plastic zone area for 2.3, 2.5, and 2.8 coordination samples with overlaid map of structural features. (b) Sample bridges for lengths of 1, 2, 3, and 4 mm (characteristic lengths). (c) Distributions of bridge lengths for 2.3,

2.5, and 2.8 coordination samples. (d) Samples with highlighted structural features ahead of the crack tip that contribute to the plastic zone shape for 2.3, 2.5, and 2.8 coordination samples. (e) Sample structural features with their constitutive bridges for 2.3, 2.5, and 2.8 coordination samples (gray, green, red, respectively). (f) Distributions of angles of alignment of structural features for 2.3, 2.5, and 2.8 coordination samples.

### 3.4. Plastic Zone Tailoring

With the two critical length scales in mind (the size of the plastic zone that is able to develop, and the population of structural features composing a given coordination number), we then seek to control the fracture response of the composite materials by spatially tailoring the coordination number. We create laminated assemblies with 25% high coordination and 75% low coordination regions in the sample (high-R edge) and then the inverse, with 25% low coordination and 75% high coordination regions (low-R edge) and perform the same single edge notch plate tension tests (Figure 5a). We observe that the low-R edge makes it possible for the development of a significantly larger plastic zone in the 2.8 coordination region than the 100% 2.8 coordination samples previously discussed (Figure 5b,c). In contrast, the high-R edge prevents the formation of as large of a plastic zone in the 2.3 coordination region as observed in the 100% 2.3 coordination samples previously discussed (Figure 5b,d). This 'flipping' effect is likely the result of the similarity in the matrix and reinforcing phase stiffnesses, which allows the entire structure to simultaneously engage in the deformation, prior to and during plastic zone formation. It is also important to note the effect of the interface where the low coordination region transitions to the high coordination region, and vice versa, which is only one characteristic length wide. This transition line causes the vertical flattening of the plastic zone in the high-R edge sample, as the population of structural features abruptly transitions from larger and more extensible to smaller and less extensible (Figure 5d). In contrast, the low-R edge sample transitions from smaller structural features to larger structural features, allowing for greater extensibility and a larger plastic zone ahead of the crack (Figure 5c).

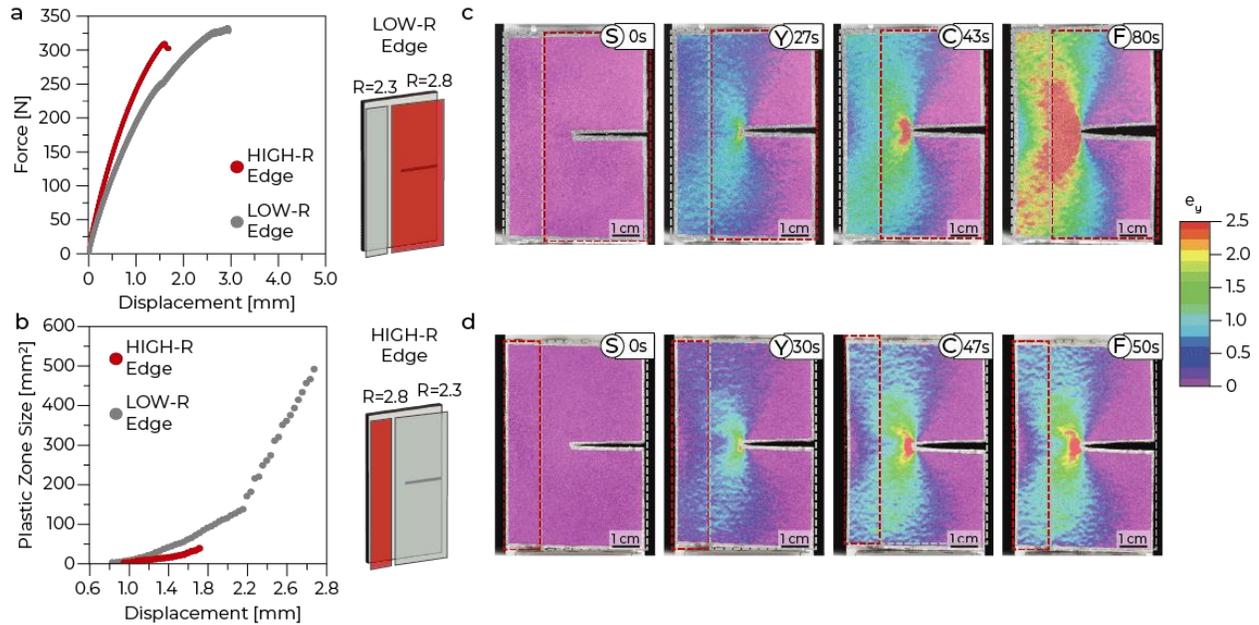

**Figure 5.** Tailoring the plastic zone. (a) 2.3 coordination and 2.8 coordination edge sample force-displacement plot. (b) Plastic zone size growth for 2.3 and 2.8 coordination edge sample. (c) 2D DIC strain maps for 2.3 coordination edge sample. (d) 2D DIC strain maps for 2.8 coordination edge sample.

## 4. Conclusions

We study the fracture behavior of nearly isodense bioinspired polymer composites with irregular network reinforcing phases. The composites feature a matrix phase with a similar stiffness but half the strength of the reinforcing phase, allowing for the formation and extension of a primary crack and therefore J-integral and R-curve measurements. We compare the effect of coordination number, a global scale descriptor, to the mechanical properties of the reinforcing phase, through measurements of plastic zone size and crack extension. We observe that low coordination samples dissipate fracture energy through the formation of a large plastic zone, and fail at higher global strain, while high coordination samples dissipate energy through crack extension, and fail at lower global strain. We determine that there are two critical length scales that explain the fracture response of the polymer composites and the variations across the coordination numbers: (i) the size of the plastic zone, and (ii) the size and geometry of the structural features. Finally, we discuss how to tailor the tradeoff between plastic zone size and primary crack extension using the coordination number and its effect on the structural feature populations ahead of the crack tip.

## Acknowledgements

The authors acknowledge F. Bouville, V. Vitelli, and D. Kammer for fruitful discussions. The authors thank P. Arakelian for the experimental assistance. C.F. and C.D. acknowledge MURI ARO W911NF-22-2-0109 for the financial support.

**Supporting Information of:**

**Fracture Characterization of Bioinspired Irregular Network Reinforced Composites**

*Chelsea Fox[1], Tommaso Magrini[2*] and Chiara Daraio[1*]*

[1] Division of Engineering and Applied Science, California Institute of Technology, Pasadena, CA 91125, USA

[2] Department of Mechanical Engineering, Eindhoven University of Technology, 5600MB Eindhoven, The Netherlands

* Email: t.magrini@tue.nl, daraio@caltech.edu

**Legend**

Supporting Information Figures: Figures S1- S5

Supporting Information Discussion 1

**Supporting Information Figures**

**Figure S1**

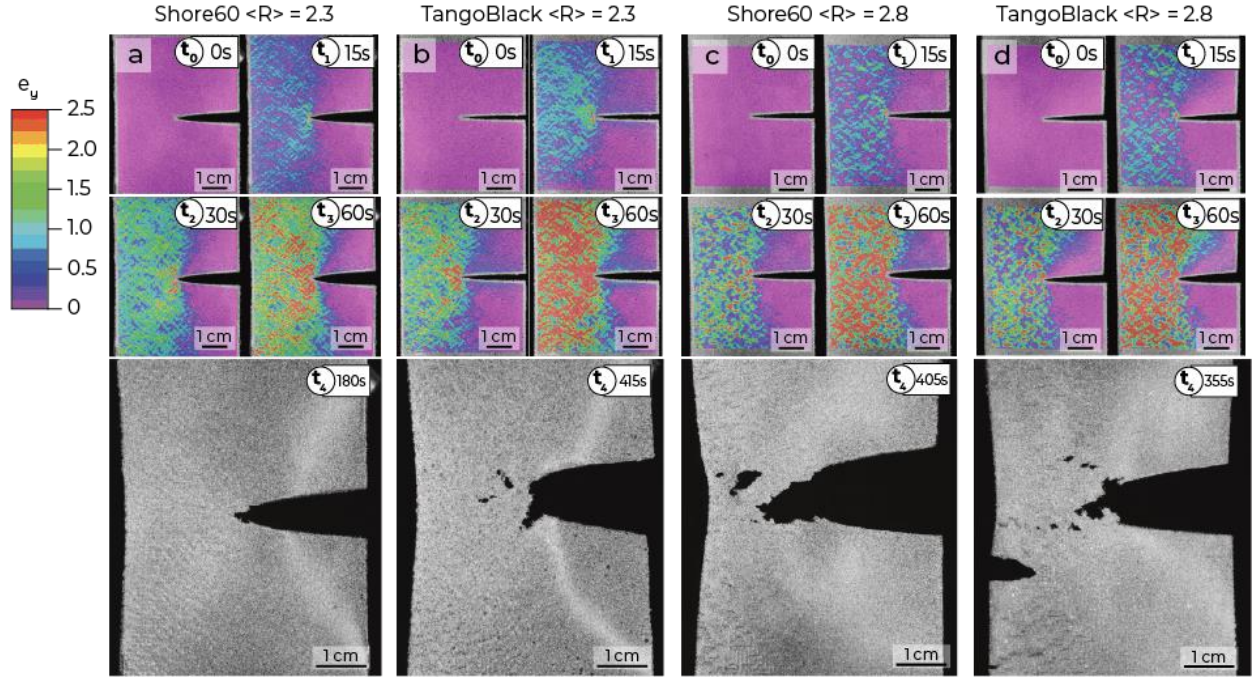

Figure S1. DIC images and fracture nucleation images for composite samples with different matrix phases. (a) 2D DIC strain maps and fracture nucleation image for 2.3 coordination composite sample with Shore60 matrix. (b) 2D DIC strain maps and fracture nucleation image for 2.3 coordination composite sample with TangoBlack matrix. (c) 2D DIC strain maps and fracture nucleation image for 2.8 coordination composite sample with Shore60 matrix. (d) 2D DIC strain maps and fracture nucleation image for 2.8 coordination composite sample with TangoBlack matrix.

**Figure S2**

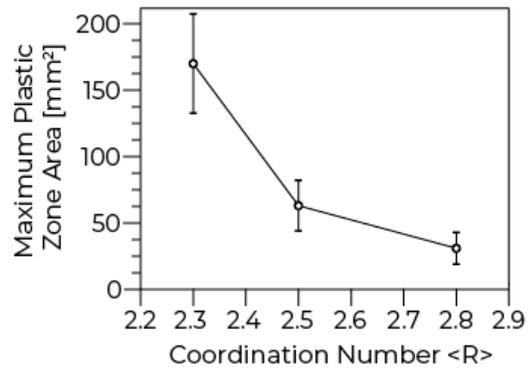

Figure S2. Maximum plastic zone size reached as a function of coordination number for 2.3, 2.5, and 2.8 samples.

**Figure S3**

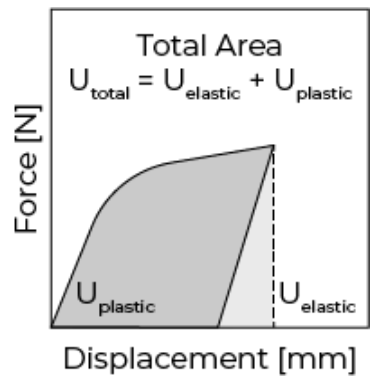

Figure S3. Calculating the incremental J-integral as elastic and plastic components.

**Figure S4**

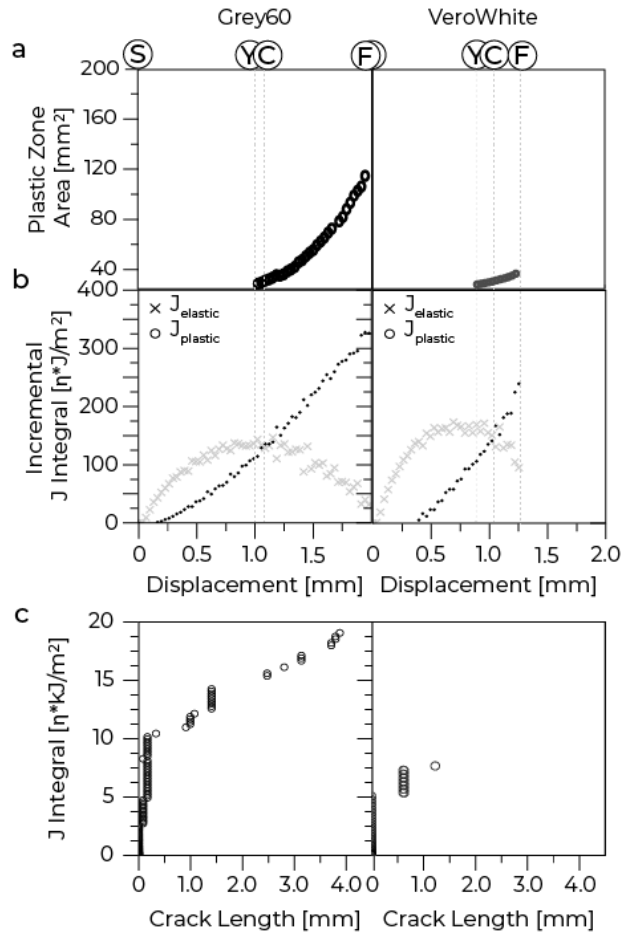

Figure S4. Plastic zone size, J integral and R-curves for bulk material samples. (a) Plastic zone area as a function of displacement for Grey60 and VeroWhite samples. S corresponds to start of the test, Y corresponds to time at yield, C corresponds to time at crack initiation, F corresponds to time at failure. (b) Incremental elastic J integrals ($\eta_e$ normalized) and plastic J integrals ($\eta_p$ normalized) as a function of displacement for Grey60 and VeroWhite samples. (c) R-curves for Grey60 and VeroWhite samples.

**Figure S5**

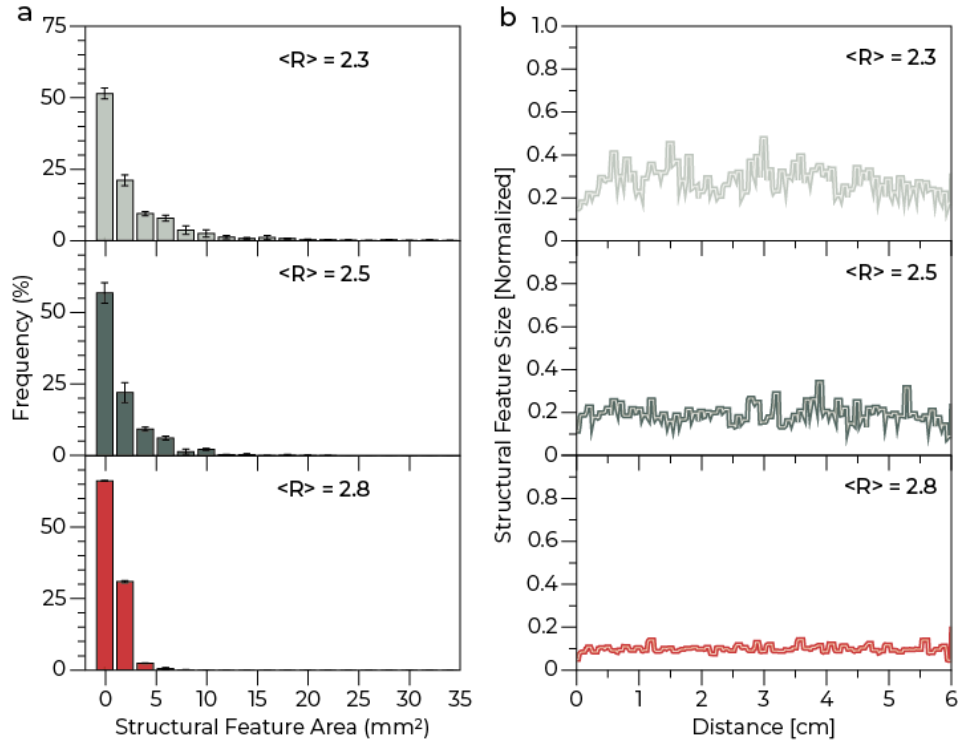

Figure S5. Structural feature characterization. (a) Structural feature area distributions for 2.3 coordination (top), 2.5 coordination (middle) and 2.8 coordination (bottom) samples. (b) Structural feature area profiles as a function of distance across the sample for 2.3 coordination (top), 2.5 coordination (middle) and 2.8 coordination (bottom) samples.

**Supporting Information Discussion 1 | Structural Feature Characterization**

By plotting the profile of feature sizes across the samples, we also observe that features are isotropically distributed, and that the average size profile varies across the coordination numbers (Figure SI 5). For the 2.8 coordination samples, the smallest features appear as a high frequency with a minimum wavelength of one characteristic length, while for the 2.5 and 2.8 coordination samples, which can form larger features, lower frequencies emerge in addition to the higher frequencies, all of which are superimposed as a result of averaging across the entire structure (Figure SI 5).